\documentclass[aps,prl,twocolumn,amsmath,longbibliography]{revtex4-2}
\usepackage{graphicx}% Include figure files
\usepackage{subfigure}
\usepackage{url}
\usepackage{dcolumn}% Align table columns on decimal point
\usepackage{bm}% bold math
\usepackage[]{natbib}
\usepackage{bbm}
\usepackage{mathrsfs}
\usepackage{amsmath}
\usepackage{amsfonts}
\usepackage[colorlinks=true, 
citecolor=blue, 
anchorcolor=blue, 
linkcolor=blue, 
urlcolor=blue]{hyperref}
\usepackage{graphicx,epstopdf}
\usepackage{subfigure}
\usepackage{epsfig}
\usepackage{dcolumn}
\usepackage{bm}
\usepackage{color}
\usepackage{natbib}
\usepackage{amssymb}
\usepackage{xcolor}
\usepackage{braket}
\begin{document}

\title{\emph{Ab Initio} Theory of Phonon Magnetic Moment Induced by Electron-Phonon Coupling in Magnetic Materials}

\author{Fuyi Wang$^{1}$, Xinqi Liu$^{1}$, Hong Sun$^{2}$, Huaiqiang Wang$^{2,5,\ast}$, Shuichi Murakami$^3$, Lifa Zhang$^2$, Haijun Zhang$^{1,4,5,\ast}$ and Dingyu Xing$^{1,4,5}$}

\affiliation{
$^1$ National Laboratory of Solid State Microstructures, School of Physics, Nanjing University, Nanjing 210093, China\\
$^2$ Center for Quantum Transport and Thermal Energy Science, Institute of Physics Frontiers and Interdisciplinary Sciences, School of Physics and Technology, Nanjing Normal University, Nanjing 210023,China\\
$^3$ Department of Physics, Tokyo Institute of Technology, 2-12-1 Ookayama, Meguro-ku, Tokyo 152-8551, Japan\\
$^4$ Collaborative Innovation Center of Advanced Microstructures, Nanjing University, Nanjing 210093, China\\
$^5$ Jiangsu Physical Science Research Center, Nanjing 210093, China}

\email{zhanghj@nju.edu.cn}
\email{hqwang@njnu.edu.cn}

\date{\today}

\begin{abstract}

Circularly polarized phonons, characterized by nonzero angular momenta and magnetic moments, have attracted extensive attention. However, a long-standing critical issue in this field is the lack of an approach to accurately calculate phonon magnetic moments resulting from electron-phonon coupling (EPC) in realistic materials. Here, based on the linear response framework, we develop an \emph{ab initio} theory for calculating EPC-induced magnetic properties of phonons, applicable to both insulating and metallic materials. Our method can precisely calculate phonon Zeeman splittings in magnetic metals with significant EPC, as demonstrated by the remarkable agreement with recent experimental observations of phonon Zeeman splitting in the ferromagnetic Weyl semimetal $\mathrm{Co_3Sn_2S_2}$. In addition, the long-sought magnetic phonon spectra across the entire Brillouin zone are obtained, facilitating the study of magnetic phonon transport and topology. Specifically, by constructing an inertially decoupled lattice model, we propose candidate materials exhibiting intrinsic phonon Chern states with robust unidirectional edge phonon currents. Our work paves the way for investigating novel phonon phenomena in magnetic quantum materials.

\end{abstract}

\maketitle

\emph{Introduction.}
Circularly polarized phonons, collective vibrations stemming from the circular motion of atoms in solids that carry nonzero angular momentum~\cite{zhang_angular_2014, zhang_chiral_2015}, are referred to as axial phonons~\cite{juraschek_chiral_2025}. Recently, there has been a surge of research interest in circularly polarized phonons~\cite{zhang_chirality_2025,zhang_new_2025}, covering both theoretical investigations~\cite{chen_chiral_2018,juraschek_orbital_2019, chen_propagating_2021,juraschek_giant_2022, saparov_lattice_2022,fransson_chiral_2023,geilhufe_electron_2023,bonini_frequency_2023,ren_phonon_2021,xiao_adiabatically_2021, zhang_gate-tunable_2023,zhang_chiral_2022,ren_adiabatic_2024, chen_gauge_2024,chaudhary_giant_2024,shabala_phonon_2024,merlin_magnetophononics_2024,xue_extrinsic_2025,kahana_light_2024,klebl_ultrafast_2025,paiva_dynamically_2025,mustafa_origin_2025} and experiments~\cite{zhu_observation_2018, cheng_large_2020,tauchert_polarized_2022, xiong_effective_2022, luo_large_2023, nabi_accurate_2023, ueda_chiral_2023, hernandez_chiral_2023, lujan_spinorbit_2024,wu_magnetic_2025,yang_inherent_2025,tang_exciton_2024}. Of particular importance is the phonon magnetic moment (PMM), which is derived from phonon energy shift in a magnetic field~\cite{ren_phonon_2021}. PMM plays a crucial role in various effects involving circularly polarized phonons, such as ultrafast Einstein-de Haas Effect~\cite{tauchert_polarized_2022} and Barnett effect~\cite{luo_large_2023,davies_phononic_2024}. However, classical models based on point-charge approximations yield PMM values that are orders of magnitude below experimental measurements~\cite{zhang_gate-tunable_2023,cheng_large_2020}. While electron-phonon coupling (EPC) has been indicated as a promising route to enhance PMMs~\cite{ren_phonon_2021,zhang_gate-tunable_2023}, to our knowledge, there is no comprehensive \emph{ab initio} framework to quantitatively predict EPC-induced PMMs in realistic materials. Previous theories describing EPC-induced PMM are only applicable to either insulators or semimetals~\cite{ren_phonon_2021,xiao_adiabatically_2021,zhang_gate-tunable_2023,bonini_frequency_2023,chen_gauge_2024,shabala_phonon_2024,merlin_magnetophononics_2024,chaudhary_giant_2024}, and more importantly, most of them rely on empirical parameters and, consequently, deviate from \emph{ab initio} approach.

%However, in metallic systems, EPC effects are more pronounced, thus the theoretical limitations have significantly constrained progress in this field.

\begin{figure}[htp]
	\includegraphics[width=0.45\textwidth]{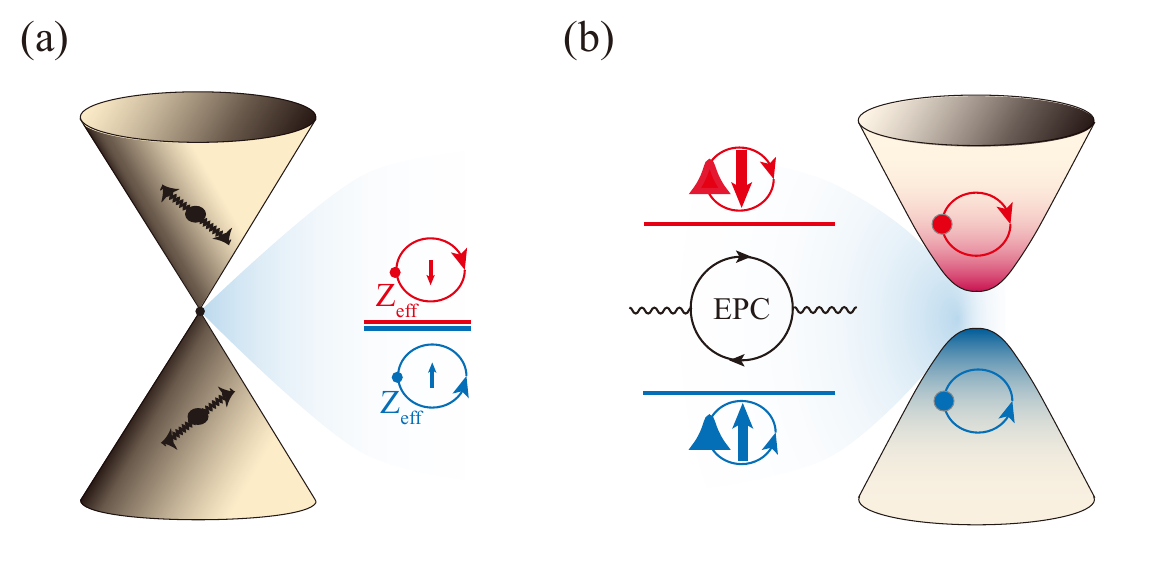}
	\caption{\label{fig:f1}Schematic of the phonon Zeeman splitting driven by (a) the classical PMM and (b) the EPC-enhanced PMM. In (a), the phonon band crossing point consists of two degenerate phonon modes vibrating in two different directions (black arrows), which can be recombined into left-circularly (red) and right-circularly (blue) polarized phonons. The classical phonon Zeeman splitting deduced from the point-charge model ($Z_{\text{eff}}$ is the effective charge) is negligible as a result of the significant atomic mass, whereas the EPC-induced phonon Zeeman splitting (b) is orders of magnitude larger. }
\end{figure}

Meanwhile, the coupling between PMM and magnetic order through spin-phonon interactions yields rich phenomena, such as phonon Zeeman splittings, phonon Hall effects~\cite{strohm_phenomenological_2005,sheng_theory_2006,kagan_anomalous_2008,Sun2020}, and magnetic topological phonon states~\cite{liu_model_2017, zhang_new_2025}. For example, phonon Zeeman splitting induced by magnetic order was recently observed experimentally in the ferromagnetic Weyl semimetal $\mathrm{Co_3 Sn_2 S_2}$~\cite{che_magnetic_2025}, yet a theory capable of quantitatively characterizing such Zeeman splittings is lacking. Moreover, investigations of phonon transport and phonon topology in magnetic systems usually require the full notion of the magnetic phonon spectra across the entire Brillouin zone. A prominent example is the phonon Chern state (phonon analogue of quantum anomalous Hall effect), which necessitates integrating Berry curvatures throughout the Brillouin zone to correctly obtain the topological Chern number~\cite{zhang_topological_2010, qin_berry_2012, liu2018berry, liu_pseudospins_2017, park_phonon_2020}. Although nonmagnetic phonon topology has been comprehensively investigated~\cite{susstrunk_classification_2016, peng_topological_2020, li_computation_2021, xu_catalog_2024, zhang_unconventional_2023, ding_topological_2024}, magnetic phonon topology remains largely unexplored due to challenges of calculating magnetic phonon spectra.

To address these challenges, we establish an \emph{ab initio} theory for determining EPC-induced magnetic properties of phonons, which enables \emph{ab initio} calculations of previously inaccessible magnetic phonon spectra across material classes from insulators to metals. Based on this theory, we demonstrate that EPC-induced PMMs can lead to large phonon Zeeman splittings in magnetic metals [Fig.~\ref{fig:f1}(b)], which are orders of magnitude stronger than classical predictions from point-charge models [Fig.~\ref{fig:f1}(a)]. Notably, our theory is further validated by the remarkable agreement with the experimental result of the phonon Zeeman splitting of $\mathrm{Co_3 Sn_2 S_2}$.
%Notably, our theoretical prediction of the phonon Zeeman splitting of $\mathrm{Co_3 Sn_2 S_2}$ shows remarkable agreement with the experimental result.} 
By applying our theory, we predict a family of promising candidate materials with phonon Chern states according to the proposed inertially decoupled lattice model. Our work establishes an \emph{ab initio} theoretical foundation for investigating EPC-induced magnetic properties of phonons, which paves the way for further developments of phonon physics.

\emph{Ab initio theory for phonon Zeeman energy and phonon magnetic moment.} Previous theories addressing magnetic properties of phonons driven by EPC have mainly been applicable to insulators or semimetals, and are restricted to the $\Gamma$ point~\cite{ren_phonon_2021,zhang_gate-tunable_2023,bonini_frequency_2023,chen_gauge_2024}. Since EPC is typically much more significant in metallic systems, we develop an \emph{ab initio} theory that can generally calculate EPC-induced magnetic properties of phonons across the whole Brillouin zone for both metals and insulators, which is also compatible with previous theories [see the supplemental material (SM)~\cite{supp}]. We will demonstrate our theory as follows.

Instead of directly calculating the magnetic moment of a single phonon, we focus on experimentally observable phonon Zeeman energy induced by the coupling between PMM and spins of electrons. We start from the total phonon Hamiltonian including the time-reversal symmetry (TRS)-breaking perturbation as
\begin{equation}
  H=\sum_{l,a}\frac{1}{2}p_{l,a}^2+\frac{1}{2}\sum_{l,ab} D_{l,a b} u_{l,a} u_{l,b}-\sum_{l,ab}\eta_{l,a b}p_{l,a} u_{l,b},
\label{eq:H1}
\end{equation}  
where $a,b=x,y,z$ represent the Cartesian coordinates, $l=1,2,... N$ denotes the lattice site for a system containing $N$ atoms, and $u_{l,a}=\sqrt{m}x_{l,a}$ represents the renormalized atomic displacement with $m$ being the atomic mass. The last term in Eq.~(\ref{eq:H1}), denoted as $\delta H$, describes the TRS-breaking perturbation resulting from spin-phonon interaction~\cite{zhang_angular_2014}, and $\eta_{l,ab}$ is a $3\times 3$ real antisymmetric matrix satisfying $\eta_{l,ab}=-\eta_{l,ba}$. In addition,
$p_{l,a}=\dot{u}_{l,a}+\sum_{l,b}\eta_{l,ab}u_{l,b}$ represents the canonical momentum~\cite{supp}. By identifying the phonon angular momentum as $s^{\text{ph}}_{l,c}=\sum_{ab}\epsilon_{abc}u_{l,a} p_{l,b}$ (c is the direction of angular momentum), with $\epsilon_{abc}$ being the third-order Levi-Civita tensor, we can express $\delta H$ as phonon Zeeman energy in a compact form $E_{\textrm{Z}}=\bm{\eta}\cdot\bm{s}^{\text{ph}}=\sum_{l,ab}\eta_{l,c}s^{\text{ph}}_{l,c}$, where we have defined $\eta_{l,c}=\frac{1}{2}\sum_{ab}\epsilon_{abc}\eta_{l,ab}$. Notably, we find that phonon angular momentum can be treated as an effective perturbative field to the electron-phonon coupled system, and thus phonon Zeeman energy can be effectively regarded as a linear response. 

Importantly, we will show that $\bm{\eta}$ can be derived from the TRS-breaking self-energy correction of phonons. 
According to $E_{\textrm{Z}}=\bm{\eta}\cdot\bm{s}^{\text{ph}}$, we have
\begin{equation}
%    \eta_{l,c}=\frac{\partial H}{\partial s^{\text{ph}}_{l,c}}=\frac{\partial H}{\partial u_{l,a}}/(\frac{\partial s^{\text{ph}}_{l,c}}{\partial u_{l,a}})=\frac{1}{p_{l,b}}\frac{\partial H}{\partial u_{l,a}}
    \eta_{l,c}=\frac{\partial H}{\partial s^{\text{ph}}_{l,c}}=\frac{1}{p_{l,b}}\frac{\partial H}{\partial u_{l,a}}
\end{equation}
where we have applied the chain derivative rule.
Applying Kubo formula, we can get
\begin{equation}
    \left<\eta_{l,c}\right>=-\frac{i}{\hbar}\int_{-\infty}^t dt' u_{l,a}p_{l,b} \left\langle\left[\frac{-1}{p_{l,a}}\frac{\partial H}{\partial u_{l,b}}(t),\frac{1}{p_{l,b}}\frac{\partial H}{\partial u_{l,a}}(t')\right]\right\rangle _0. 
    \label{eq:eta0}
\end{equation}
By taking Fourier transform of Eq.~(\ref{eq:eta0}) and considering $p_{l,a}=-i\omega u_{l,a}$, we can obtain $\eta^c(\boldsymbol{q}, \omega)$ ($\boldsymbol{q}$ is the phonon wave vector, and $\omega$ denotes phonon frequency) as
\begin{equation}
        \eta_{c}(\boldsymbol{q}, \omega)=\frac{1}{\hbar\omega} \int d t e^{-i \omega t}\theta(t)\left\langle\left[\hat{V}_{a}(\boldsymbol{q},\omega), \hat{V}_{b}(-\boldsymbol{q},\omega)\right]\right\rangle _0.
\end{equation}
Here, $\theta(t)$ is the Heaviside step function, and the EPC interaction operator $
\hat{V}_{a}(\boldsymbol{q},\omega)=\sum_{l_c}z_{l_c}(\boldsymbol{q},\omega)\frac{\partial H}{\partial u_{l,a}}$, where $z_{l_c}(\boldsymbol{q},\omega)$ is the normalized contribution factor of the $l_c$-th atom for the phonon mode $(\boldsymbol{q},\omega)$\cite{supp}.

Now, we take electron's spin degree of freedom into consideration.
Taking the spin-quantization axis of electron as the c direction, we could get the relation that $\bm{\eta}(\bm{q},\omega)=\frac{1}{i\omega}\bm{\chi}^{-}(\bm{q},\omega)$, with $\chi_{c}^{-}(\bm{q},\omega)=\chi_{c,\uparrow}(\bm{q},\omega)-\chi_{c,\downarrow}(\bm{q},\omega)$ (see SM~\cite{supp} for details), indicating that the spin polarization of electrons leads to phonon Zeeman splittings.
The susceptibility $\chi_{c,s}(\bm{q},\omega)$ for $s=\uparrow,\downarrow$ can be obtained by performing the Matsubara sum, 
\begin{equation}
    \chi_{c,s}(\bm{q},\omega)=\frac{2i\omega}{\hbar N_{\bm{k}}}\sum_{\bm{k} ,\alpha\beta}\left|{V_{c,s}^{\bm{k},\alpha\beta}}(\bm{q},\omega)\right|^2\frac{f(\epsilon_{s}^{\bm{k}+\bm{q},\alpha})-f(\epsilon_{s}^{\bm{k},\beta})}{\hbar\omega-(\epsilon_{s}^{\bm{k}+\bm{q},\alpha}-\epsilon_{s}^{\bm{k},\beta})}.
    \label{eq:chi}
\end{equation}
Here, $\bm{k}$ represents the wave vector of the electron, $N_{\bm{k}}$ is the total number of $\bm{k}$, and $\epsilon_{s}^{\bm{k},\alpha}$ denotes the electronic energy with band indices $\alpha$. $f$ is the Fermi-Dirac distribution, and $V_{c,s}^{\bm{k},\alpha\beta}(\bm{q},\omega)$ is the EPC tensor~\cite{giustino_electron-phonon_2017,zhou_ab_2021}. It should be emphasized that all parameters in Eq.~(\ref{eq:chi}) can be directly obtained from \emph{ab initio} calculations, without empirical inputs or additional assumptions. A non-zero $\bm{\chi}^{-}$ requires spin-polarized electronic states, for example, in ferromagnets. To realize large $\bm{\chi}^{-}$, there are two key points according to Eq.~(\ref{eq:chi}): (i) The contribution of EPC tensor $V_{c,s}^{\bm{k},\alpha\beta}(\bm{q},\omega)$ is substantial, and materials composed of light atoms tend to possess strong $V_{c,s}^{\bm{k},\alpha\beta}(\bm{q},\omega)$. (ii) When the energy difference $(\epsilon^{\bm{k}+\bm{q},\alpha}-\epsilon^{\bm{k},\beta})$ approaches $\hbar \omega$, it leads to an enhancement of $\bm{\chi}^{-}$. Consequently, we can conclude that EPC-induced large phonon Zeeman splittings are most likely to occur in ferromagnetic narrow-gap semiconductors, semimetals, and metals containing light elements. 

Then, PMM can be derived from the above phonon Zeeman energy as (see SM~\cite{supp})
\begin{equation}
  \mu^{\text{ph}}_{c}=-\frac{\partial E_{\text{Z}}}{\partial B_c}=-g^{\text{ph}}_{c}\frac{\mu_{B}}{\hbar}s^{\text{ph}}_{c}.
  \label{eq:pmm}
\end{equation}
Here, $\mu_B$ denotes the Bohr magneton of electron, $B_c$ is the effective magnetic field, and $g^{\text{ph}}_{c}=g^{e}\frac{\hbar}{2i\omega}\frac{\partial \chi_c^{+}}{\partial E}|_{E=E_F} $ represents the effective phonon $g$-factor   (see SM~\cite{supp}), where $\chi_{c}^{+}(\bm{q},\omega)=\chi_{c,\uparrow}(\bm{q},\omega)+\chi_{c,\downarrow}(\bm{q},\omega)$ with $g^{e}$ being the electronic $g$-factor involved in EPC. Equation~(\ref{eq:pmm}) enables \emph{ab initio} calculations of PMM. The dimensionless parameter $\frac{\hbar}{2i\omega}\frac{\partial \chi_c^{+}}{\partial E}|_{E=E_F}$ acts as an amplification factor dominated by the EPC tensor $V_{c,s}^{\bm{k},\alpha\beta}$ and density of electron states near the Fermi energy [Eq.~(S28) in SM~\cite{supp}], and it allows the $g$-factor of the circularly polarized phonon to reach the order of $g^{e}$, leading to a large PMM. This enhancement of PMM originates from electron loop currents induced by the EPC-mediated angular momentum transfer from ions to electrons, where electron loop currents are significantly enhanced owing to the tiny electron mass. In contrast, within the classical framework, the phonon's $g$-factor is constrained to $10^{-5}\sim10^{-4}$ order of $g^{e}$ due to large ionic masses compared to the electron mass~\cite{zhang_gate-tunable_2023,wang_chiral_2022}. 

Based on the above theory, we develop a dedicated computational module within Quantum ESPRESSO~\cite{giannozzi_quantum_2009,giannozzi_advanced_2017} incorporating density-functional perturbation theory with Hubbard corrections (DFPT+U)~\cite{giustino_electron-phonon_2017,zhou_ab_2021} for precise determination of response coefficients $\bm{\eta}$ and Landé $g$-factors of phonons. Then we add the effect of $\bm{\eta}$ into Schrödinger-like phonon equations constructed from the dynamical matrix through near-degenerate perturbation theory (see SM~\cite{supp} for details\nocite{baroni_phonons_2001,zhang_deep_2022,togo_implementation_2023,togo_first-principles_2023,fukui_chern_2005,jain_commentary_2013,gga}), yielding TRS-breaking phonon dispersions. 

\begin{figure}
  \includegraphics[width=0.5\textwidth]{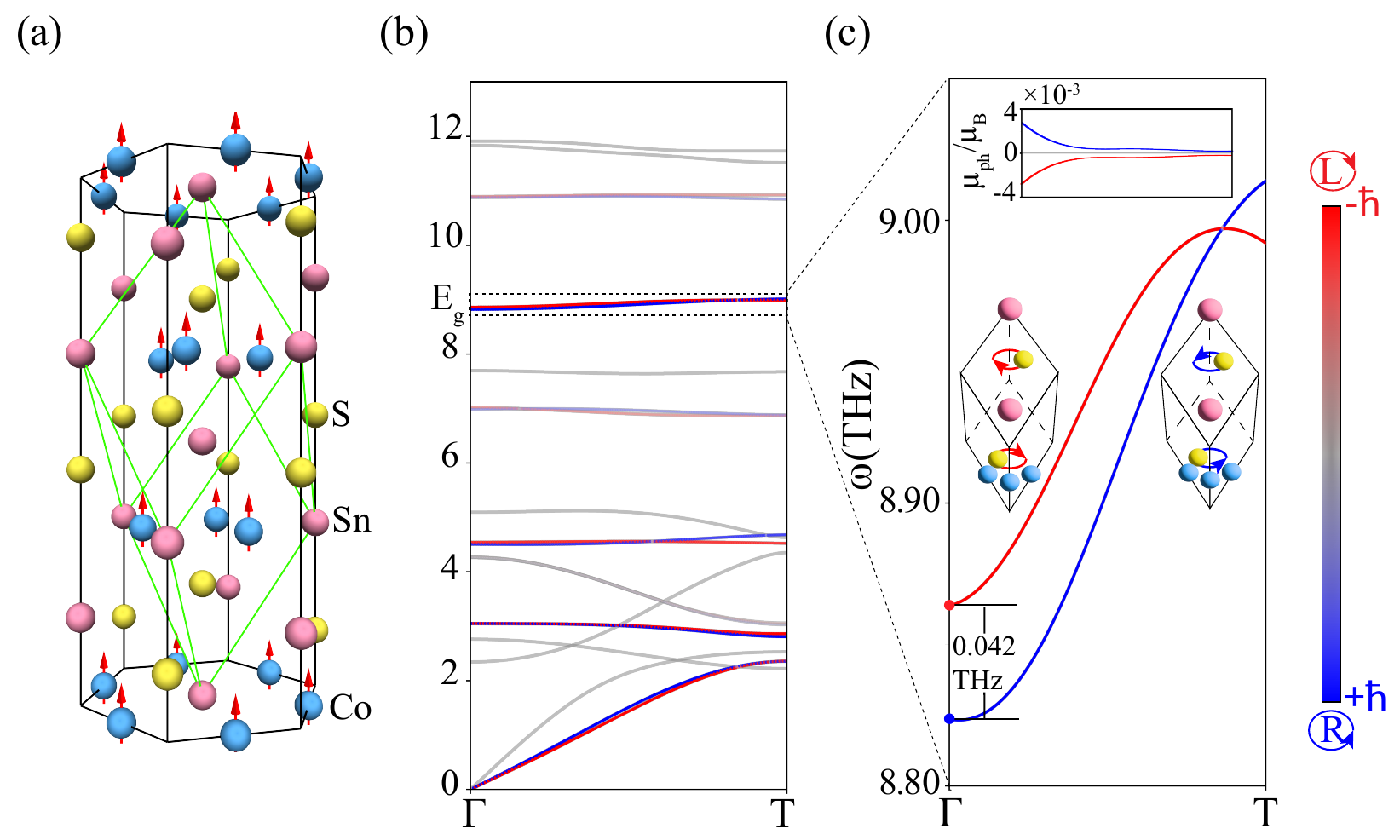}
    \caption{\label{fig:f_x} Lattice structure and calculated magnetic phonon dispersions of $\mathrm{Co_3Sn_2S_2}$. (a) The conventional cell (black lines) and primitive cell (green lines) of $\mathrm{Co_3Sn_2S_2}$. (b) The magnetic phonon dispersion with phonon Zeeman splittings along $\Gamma$-T path. (c) The magnetic phonon dispersion of $E_g$ modes. The color gradient from red to blue represents the phonon angular momentum. Upper insets: The phonon magnetic moment for the two branches of $E_g$ mode along $\Gamma$-T path.} Lower insets:  Schematic of the two phonon modes with opposite angular momenta, corresponding to right-handed and left-handed circular motions of S atoms, respectively.
\end{figure}

\emph{Comparison with experiment.} A recent experimental work reported the first  observation of magnetic order induced circularly polarized phonons and their spontaneous splitting in the ferromagnetic Weyl semimetal $\mathrm{Co_3Sn_2S_2}$~\cite{che_magnetic_2025}. As shown in Fig.~\ref{fig:f_x}(a), $\mathrm{Co_3Sn_2S_2}$ belongs to centrosymmetric space group $\mathrm{R\bar{3}m}$, and exhibits out-of-plane ferromagnetic ordering below the Curie temperature ($\sim$175 K). Using helicity-resolved magneto-Raman scattering spectroscopy, a discernible splitting of the $E_g$ phonon modes at the $\Gamma$ point was observed in the ferromagnetic state, reaching up to $1.27~\mathrm{cm}^{-1} $($\sim$0.038 THz) at low temperatures (2 K). The two split modes correspond to phonon modes with opposite angular momenta, describing the right-handed and left-handed in-plane circular motion of S atoms, respectively [see Fig.~\ref{fig:f_x}(c) insets]. Although Ref.~\cite{che_magnetic_2025} provides an implicit EPC-based explanation of such a notable phonon Zeeman splitting, a direct and more convincing quantitative description is absent due to the lack of \emph{ab initio} methods for magnetic phonon spectra.

Now we show that our \emph{ab initio} method could well reproduce the experimental result quantitatively. Giving full consideration to EPC effects, for the first time, we obtain the complete magnetic phonon dispersions of $\mathrm{Co_3Sn_2S_2}$ throughout the Brillouin zone (see the SM for more details~\cite{supp}). Figure ~\ref{fig:f_x}(b) presents the experimentally relevant magnetic phonon dispersion along the high-symmetry path $\Gamma$-T. Remarkably, our calculations clearly demonstrate a splitting of 0.042 THz ($\mu_\text{ph}=2.8\times 10^{-3} \mu_B$) at the $\Gamma$ point for the $E_g$ mode [Fig.~\ref{fig:f_x}(c)], in good agreement with the observed value of 0.038 THz. Beyond this, we surprisingly  find that the two circularly polarized phonon modes cross each other on the $\Gamma$-T path, forming a phonon Weyl point [Fig.~\ref{fig:f_x}(c)]. This magnetic phonon Weyl point \cite{zhang_weyl_2025} is protected by the $C_{3v}$ point group symmetry along $\Gamma$-T, which is previously inaccessible without obtaining the full magnetic phonon dispersions.

\emph{Intrinsic phonon Chern states and the inertially decoupled model.}
The accurate calculation of phonon Zeeman splitting enables the determination of magnetic phonon spectra throughout the Brillouin zone, which is indispensable for exploring magnetic phonon topology. Following the analogy to the Chern state in the Haldane model~\cite{Haldane1988}, we adopt the two-dimensional (2D) honeycomb lattice~\cite{liu_model_2017,bonini_phonon_2007,li_direct_2023} as an illustrative example. When either inversion symmetry ($I$) or TRS ($T$) is broken, each phonon Dirac cone at $K$($K'$) valley acquires a finite gap, effectively gaining masses denoted as $\Delta_I$ and $\Delta_T$, respectively. It has been shown that when $|\Delta_T|>|\Delta_I|$, band inversion occurs in one of the two Dirac cones, resulting in a topological phonon state with a Chern number of $|C|=1$~\cite{liu_model_2017}.

\begin{figure}[t]
  \includegraphics[width=0.47\textwidth]{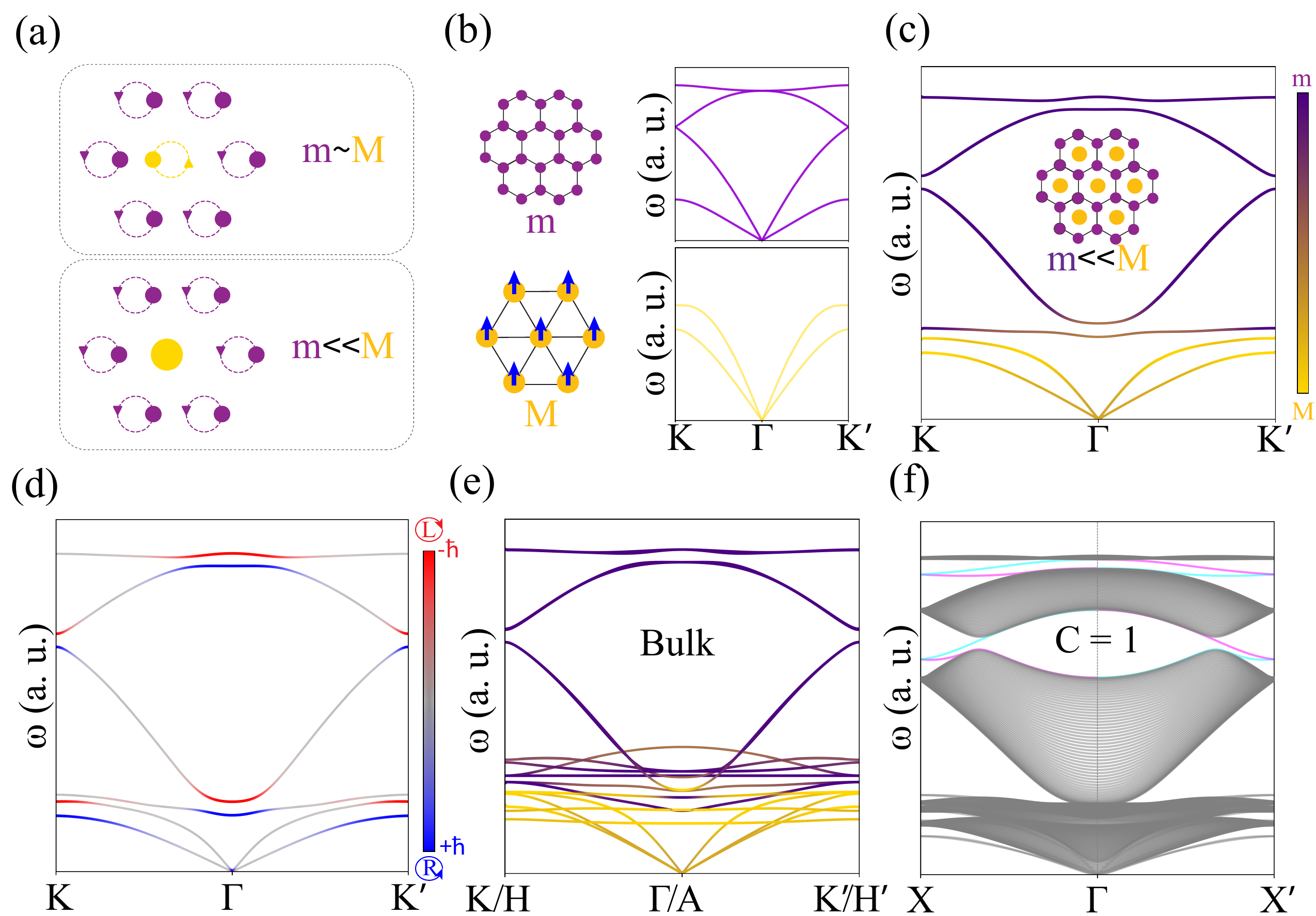}%
    \caption{\label{fig:f2}Inertially decoupled model (IDM) and intrinsic phonon Chern states. (a) Schematic of circularly polarized phonon modes in the high frequency regime: When $m\sim M$, both atoms contribute to high-frequency modes (upper panel); in the IDM condition ($m \ll M$), high-frequency modes are primarily contributed by light atoms (lower panel). (b) Phonon dispersions of nonmagnetic honeycomb lattice and ferromagnetic triangular lattice. (c-e) TRS-breaking phonon dispersions of the 2D IDM by combining the honeycomb and triangular lattices (c,d) and the 3D IDM by stacking 2D IDM along the $z$-direction (e). (f) Edge states (cyan/magenta) of the 2D IDM. Parameters are detailed in the SM~\cite{supp}.}
\end{figure} 

However, it is challenging to realize such phonon Chern states in realistic materials with honeycomb lattices. Although Coriolis forces and strong magnetic fields can produce a nonzero $\Delta_T$~\cite{liu_model_2017}, these approaches either require large external fields or complex experimental measurements. In contrast, the phonon Zeeman splitting could offer a sizable $\Delta_T$ in magnets without the need of external fields. To further satisfy the condition of $|\Delta_T|>|\Delta_I|$, it is necessary to minimize $|\Delta_I|$. Therefore, an elemental ferromagnetic honeycomb material inherently preserving inversion symmetry with $|\Delta_I|=0$ would be the ideal candidate. Unfortunately, to our knowledge, no elemental ferromagnetic honeycomb material has been discovered. %Furthermore, any honeycomb lattice composed of multiple elements could introduce a substantial $\Delta_I$, which will make the condition $|\Delta_T|>|\Delta_I|$ difficult to satisfy. 

To seek realistic materials that exhibit intrinsic phonon Chern states, we propose a practical composite lattice model, referred to as the inertially decoupled model that comprises two types of atoms with a significant mass (inertia) difference. A prototypical example of this model is constructed by combining a honeycomb lattice composed of nonmagnetic light atoms of mass $m$ with a triangular lattice of heavy magnetic atoms of mass $M$ ($m\ll M$), as shown in Figs.~\ref{fig:f2}(b,c). Due to the difference in atomic inertia, the high-frequency and low-frequency dispersions are mainly contributed by the light and heavy atoms, respectively, as shown in Figs.~\ref{fig:f2}(a,c), which is termed ``inertially decoupled'' (see SM~\cite{supp} for details). Within the honeycomb sublattice, the inversion symmetry is preserved, resulting in a vanishing $\Delta_I$ mass at both the $K$ and $K'$ valleys. Meanwhile, the magnetic $M$ atoms break TRS to generate large phonon Zeeman splittings, producing the desired large $\Delta_T$ mass. The large $\Delta_T$ opens a topologically nontrivial phonon gap (seen in Fig.~\ref{fig:f2}(d)), thereby realizing an intrinsic phonon Chern state with $|C|=1$, characterized by topologically protected edge states crossing the phonon gap, as shown in Fig.~\ref{fig:f2}(f). 
%%%%%%%%%%

The inertially decoupled model is also applicable to three-dimensional (3D) magnets. Interestingly, when the nonmagnetic honeycomb lattice (the $m$ sublayer) and the ferromagnetic triangular lattice (the $M$ sublayer) are stacked alternately along the $z$-direction, each $M$ sublayer effectively acts as a buffer layer that nearly decouples the phonon dispersions of adjacent $m$ sublayers in the high-frequency regime. A representative phonon dispersion of such a 3D bulk model is shown in Fig.~\ref{fig:f2}(e). Notably, since each $m$ sublayer realizes a Chern state with $|C|=1$, the system exhibits a 3D phonon Chern state. This configuration hosts multiple unidirectional phonon edge states resulting from a total Chern number $|C|=N_z$, where $N_z$ is the number of $m$ sublayers.

Through \emph{ab initio} calculations guided by the inertially decoupled model, we identified five hexagonal ($P6/mmm$) magnets exhibiting intrinsic phonon Chern states, namely, $\mathrm{EuSi_2}$, $\mathrm{GdSi_2}$~\cite{mayer1967dimorphism}, $\mathrm{EuGa_2}$, $\mathrm{GdGa_2}$~\cite{barandiaran1989magnetic}, and $\mathrm{CrB_2}$~\cite{grechnev_effect_2009,bauer_low-temperature_2014}. Remarkably, they all have large observable topological phonon gaps at the $\mathrm{K}$ point defined as $\Delta_{\text{ph}}=\Delta_{\text{z}}-\gamma$, where $\Delta_{\text{z}}=2|E_{\text{Z}}|$ is the phonon Zeeman splitting and $\gamma$ is the phonon linewidth resulting from EPC~\cite{debernardi_phonon_1998,tang_first-principles_2011}, as summarized in Table~I. Note that the Zeeman splitting should be larger than the phonon linewidth to produce an observable phonon gap. 

\begin{figure}[t]
  \includegraphics[width=0.45\textwidth]{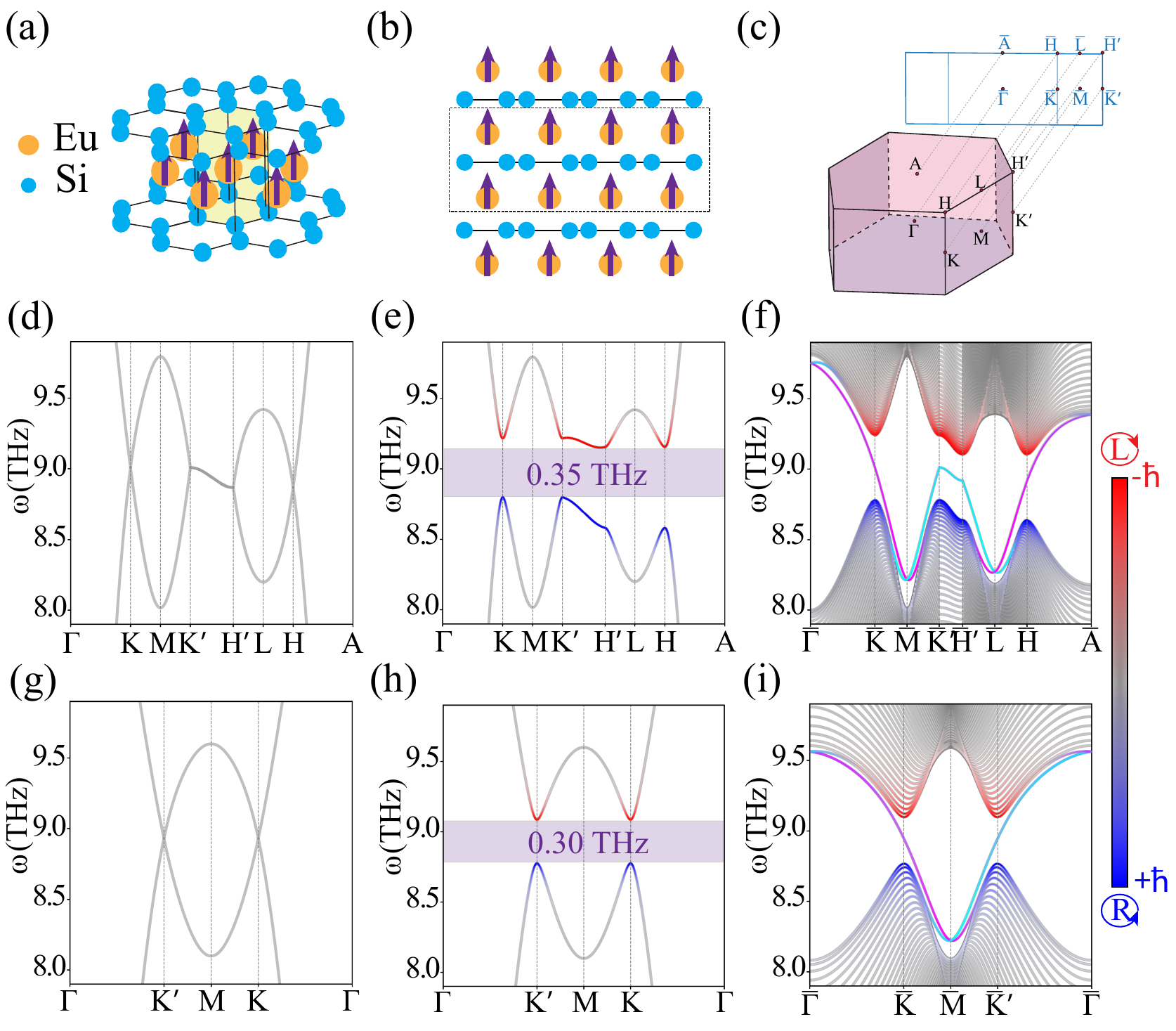}
    \caption{\label{fig:f3} Lattice structure and TRS-breaking phonon dispersions of ferromagnetic $\mathrm{EuSi_2}$. (a,b) The front view (a) and the side view (b) of the lattice structure of $\mathrm{EuSi_2}$. (c) The Brillouin zones of bulk $\mathrm{EuSi_2}$ and the projected side plane. (d,e) The phonon dispersions of bulk $\mathrm{EuSi_2}$ without (d) and with (e) the phonon Zeeman splittings, respectively. The calculated PMMs at $K(K')$ and $H(H')$ point are $5.7\times 10^{-3} \mu_B$ and $2.3\times 10^{-2} \mu_B$, respectively. (f) The phonon dispersion of bulk $\mathrm{EuSi_2}$ under open-boundary conditions.  (g,h) The phonon dispersion of the sandwiched slab enclosed by the black rectangle in (b) without (g) and with (h) the phonon Zeeman splittings, respectively. The PMM at $K(K')$ point is $2.2\times 10^{-2} \mu_B$. (i) The phonon dispersion of the sandwiched slab under open-boundary conditions. The color gradient from red to blue represents the phonon angular momentum. Cyan and magenta colors in (f) and (i) label topological boundary modes residing on opposite boundaries.}
\end{figure}

Here, we take $\mathrm{EuSi_2}$ as a representative example (see SM~\cite{supp} for details on all candidate materials). The lattice structures of $\mathrm{EuSi_2}$ are illustrated in Figs.~\ref{fig:f3}(a, b). Nonmagnetic $\mathrm{Si}$ atoms form a honeycomb-lattice layer, stacked alternately with a ferromagnetic $\mathrm{Eu}$ triangular-lattice layer, along the [001] direction. The atomic masses of $\mathrm{Si}$ and $\mathrm{Eu}$ are approximately $28\,u$ and $152\,u$ ($u$ is atomic mass unit), respectively, with each $\mathrm{Eu}$ atom carrying an out-of-plane magnetic moment of $7\,\mu_B$. As a result, $\mathrm{EuSi_2}$ can be aptly described by our proposed inertially decoupled model.

Our \emph{ab initio} calculations reveal that EPC-enhanced PMM ($\sim10^{-2} \mu_B$) induces phonon Zeeman splitting in the high-frequency phonon dispersion of ferromagnetic $\mathrm{EuSi_2}$ [see Figs.~\ref{fig:f3}(d, e)], opening a sizable full topological gap ($\approx 0.35$ THz). Given that the Chern number remains $|C|=1$ for all 2D $k_z$ slices, $\mathrm{EuSi_2}$ realizes a 3D layered phonon Chern state. This is further supported by the presence of boundary modes in the phonon dispersions of $\mathrm{EuSi_2}$ under open-boundary conditions, as illustrated in Fig.~\ref{fig:f3}(f). To achieve a 2D Chern state, we construct a dynamically-stable sandwiched slab featuring a $\mathrm{Si}$ sublayer between two $\mathrm{Eu}$ sublayers, as shown by the black rectangle in Fig.~\ref{fig:f3}(b). By comparing Figs.~\ref{fig:f3}(g) and~\ref{fig:f3}(h), we can also see that the phonon Zeeman splitting generates a topologically nontrivial phonon gap ($\approx 0.30$ THz) with a Chern number of $|C|=1$ featuring a pair of edge modes traversing the phonon gap, as shown in Fig.~\ref{fig:f3}(i). This provides definitive evidence for intrinsic phonon Chern states.

\begin{table}[htbp]
  \caption{\label{tab:table1}
  The direct phonon gaps at K points for the candidate materials, where $\gamma$ and $\Delta_{\text{z}}$ represent the phonon linewidth and the phonon Zeeman splitting, respectively. The observable phonon gap $\Delta_{\text{ph}}$ is defined as $\Delta_{\text{z}}-\gamma$.
  }
  \begin{ruledtabular}
  \begin{tabular}{cccc}
  \textrm{Mat.}&
  \textrm{$\Delta_{\text{z}}$ (THz)}&
  \textrm{$\gamma$ (THz)}&
  \textrm{$\Delta_{\text{ph}}$ (THz)}\\
  \colrule
    $\mathrm{EuSi_2}$ & 0.426 & 0.049 & 0.377\\
    $\mathrm{GdSi_2}$ & 0.406 & 0.061 & 0.345\\
    $\mathrm{EuGa_2}$ & 0.134 & 0.028 & 0.106\\
    $\mathrm{GdGa_2}$ & 0.145 & 0.015 & 0.130\\
    $\mathrm{CrB_2}$ &  0.270 & 0.033 & 0.237\\
  \end{tabular}
  \end{ruledtabular}
\end{table}

\emph{Conclusion.}
Based on the linear response theory, we have established an \emph{ab initio} theory, enabling systematic investigation of EPC-induced magnetic phonon spectra and phonon magnetic properties for a wide range of insulators, semimetals, and metals. This theoretical method is validated through the excellent agreement with the recent experiment results of magnetic Weyl semimetal $\mathrm{Co_3 Sn_2 S_2}$, providing a quantitative explanation of the phonon Zeeman splitting induced by magnetic order. We also propose an inertially decoupled model to seek intrinsic phonon Chern states in realistic materials, and predict $\mathrm{EuSi_2}$ family materials as candidates exhibiting large topological phonon gaps exceeding 0.3 THz, which could be applied to detect neutral particles, such as dark matter particles~\cite{supp}. Therefore, our work paves a broad way for investigating phonon magnetic effects and phonon topology in quantum materials.

%In summary, we have established an \emph{ab initio} theory enabling systematic predictions of EPC-induced magnetic properties of phonons and magnetic phonon dispersions for insulators, semimetals, and metals. Our theoretical method well reproduces the experiment results of $\mathrm{Co_3 Sn_2 S_2}$, and provides a quantitative explanation of the large phonon Zeeman splitting induced by magnetic order. We also propose a practical scheme to seek intrinsic phonon Chern states in realistic materials, and predict candidates exhibiting large topological phonon gaps exceeding 0.3 THz. Our work could open broad pathways for investigating phonon magnetic effects in quantum materials.

\begin{acknowledgments}
The authors thank Lei Zhang and Zuowei Liu for valuable discussions. This work is partly supported by National Key Projects for Research and Development of China (Grants No. 2024YFA1409100, No. 2021YFA1400400, No. 2022YFA1403602, and No. 2023YFA1407001), the Natural Science Foundation of Jiangsu Province (No. BK20253012, BK20252117, BK20233001, BK20243011, and BK20220032), the Natural Science Foundation of China (No. 12534007 and No. 92365203), and the e-Science Center of Collaborative Innovation Center of Advanced Microstructures. 

\end{acknowledgments}

\bibliography{references}

\end{document}